# Quadratic-nonlinear Landau-Zener Transition for Association of an Atomic Bose-Einstein Condensate with Inter-Particle Elastic Interactions Included

**A. Ishkhanyan**[1], **R. Sokhoyan**[1,2], **K.-A. Suominen**[3], **C. Leroy**[2], **and H.-R. Jauslin**[2]

[1]*Institute for Physical Research of NAS of Armenia, 0203 Ashtarak-2, Armenia*
[2]*Institut Carnot de Bourgogne, UMR 5209 CNRS, Université de Bourgogne, BP 47870, 21078 Dijon, France*
[3]*Department of Physics and Astronomy, University of Turku, 20014 Turun yliopisto, Finland*

(Dated: November 3, 2009)

**Abstract.** We study the strong coupling limit of a quadratic-nonlinear Landau-Zener problem for coherent photo- and magneto-association of cold atoms taking into account the atom-atom, atom-molecule, and molecule-molecule elastic scattering. Using an exact third-order nonlinear differential equation for the molecular state probability, we develop a variational approach which enables us to construct a highly accurate and simple analytic approximation describing the time dynamics of the coupled atom-molecule system. We show that the approximation describing time evolution of the molecular state probability can be written as a sum of two distinct terms; the first one, being a solution to a limit first-order *nonlinear* equation, effectively describes the process of the molecule formation while the second one, being a scaled solution to the *linear* Landau-Zener problem (but now with *negative* effective Landau-Zener parameter as long as the strong coupling regime is considered), corresponds to the remaining oscillations which come up when the process of molecule formation is over.

## 1. Introduction

The Landau-Zener (LZ) model [1] became long ago a standard tool in quantum physics. It describes a paradigmatic situation when two quantum states are coupled by an external field of constant amplitude and a variable frequency which is linearly changed in time. The LZ model serves as a prototype of all level-crossing models; hence, deep understanding of the LZ model will be an essential step towards intuitive perception of all level-crossing processes in general. Consequently, the LZ transition has found wide applications in many branches of physics, such as quantum control and quantum information [2-6], quantum dots [7] and molecular clusters [8], to name a few.

This particular model is one of the most used approximations in resonance physics due to its specific features. First of all, the detuning is a linear function of time, which is a realistic assumption near a resonance crossing. Second, the coupling is constant; near the crossing this is a relatively good approximation if the actual coupling changes slowly in time compared to the detuning, which is the usual situation.

It should be noted, however, that the LZ model suffers from two substantial shortcomings: the coupling does not vanish at infinite times, which implies an infinite energy (in the case of interaction with the laser field), and the detuning tends to infinity with time, which is also unphysical. Mathematically, this also leads to considerable complications compared with other models. Nevertheless, for the cases when the transitions take place in a narrow time interval around the resonance point, the time dependence of the actual coupling and the detuning far from the crossing does not considerably affect the dynamics of the system and thus the model provides an accurate description of physical processes.

When generalizing the LZ process to those associated with the mean-field dynamics of interacting many-body systems [9], one obtains nonlinear Landau-Zener (NLZ) processes for which the simple physical intuition based on linear LZ model may become invalid. The version of the NLZ problem we consider here [10-11] is a basic semiclassical variant of a non-linear two-state problem arising in nonlinear field theories involving a Hamiltonian with a 2:1 resonance [12] and time-independent quartic nonlinear terms. This model is used, e.g., in the theories of cold atom production in atomic Bose-Einstein condensates [13] via laser Raman photoassociation [14] or magnetic Feschbach resonance [15], and in the second harmonic generation in non-linear optics [16].

The basic version of the NLZ problem has been considered, e.g., in Refs. [17-28]. In these developments, not included are the quartic nonlinear terms in the Hamiltonian that, for the case of the cold molecule formation, describe inter-particle elastic scattering. One of the main conclusions one gains from the obtained results is that in the strong interaction limit the non-transition probability turns to be proportional to the inverse sweep rate, in contrast to the linear two-state case when the dependence is exponential [1]. Further, juxtaposing the results of Refs. [17-28], we see that, in contrast to the other listed works, Refs. [18-20] not only provide a prediction for the final transition probability but also suggest highly accurate analytical formulas to describe the whole temporal dynamics of the system. In particular, the absolute error of the analytical formula for the number of the associated molecules, presented in Ref. [20], does not exceed $10^{-4}$ at the end of the interaction ($t \to +\infty$) while for particular time points it may increase up to $10^{-3}$. Importantly, the mentioned formula provides the same accuracy at arbitrary values of the problem's input parameters.

The role of inter-particle interactions in the cold atom coherent association dynamics has already been discussed, e.g., in Refs. [29-32]. It has been shown that these interactions strongly affect the process of molecule formation. In particular, it has been shown that, in the case when external field configuration is defined by the LZ model, inter-particle elastic scattering is described by a sole combined parameter [32]. Moreover, it has been revealed that depending on the sign of this parameter the molecule conversion efficiency can both diminish and increase. In the present paper, by analyzing both molecule conversion efficiency and temporal dynamics of the atom association, we first define favorable conditions for formation of molecules. Further, we develop a version of the variational method [33] which not only enables one to predict the final transition probability to the molecular state but also provides a highly accurate and simple analytical formula describing the temporal dynamics of the coupled atom-molecular system for the case when the inter-particle elastic scattering is included in the basic version of the NLZ problem. The constructed analytical approximation is valid in the strong interaction limit and moderate values of the mentioned combined parameter which describes inter-particle elastic scattering. We also show that inter-particle elastic scattering results in the nonlinear shift of the effective resonance point and find an analytical expression for the effective resonance crossing time point (applicable in the strong interaction limit) written in terms of the input parameters of the problem. It should be emphasized that our approach gives an accurate analytical description of the whole temporal dynamics of the molecule formation process.

## 2. General observations

We consider the following nonlinear system of mean-field coupled Gross-Pitaevskii-type equations describing atomic and molecular condensates as classical fields [10-11]:

$$i\frac{da_1}{dt} = U(t)e^{-i\delta(t)}a_1^* a_2 + (\Lambda_{11}|a_1|^2 + \Lambda_{12}|a_2|^2)a_1,$$
$$i\frac{da_2}{dt} = \frac{U(t)}{2}e^{i\delta(t)}a_1 a_1 + (\Lambda_{21}|a_1|^2 + \Lambda_{22}|a_2|^2)a_2,$$
(1)

where $a_1$ and $a_2$ are the atomic and molecular state probability amplitudes, respectively, $t$ is the time. The detuning $\delta_t$ defines the difference in energy between a stationary molecule and two stationary atoms $\hbar\delta_t$ which can be adjusted by tuning the laser field frequency in the case of photoassociation or by variation of the magnetic field in the case of the Feshbach resonance. The function $\delta(t)$ of Eq. (1) is defined as the integral of the detuning $\delta_t$ (here, the subscript denotes the derivative with respect to time). In the case of photoassociation the atom-molecule coupling $U$ can be controlled by variation of the laser field intensity, while in the case of Feshbach resonance it is a fixed constant (we consider the case of homogeneous condensate whose density does not vary in space). In the set of equations (1), the cubic nonlinearities describe the inter-particle elastic scattering processes. The coefficients $\Lambda_{jk}$ ($j,k = 1,2$) in the diagonal case $j = k$ are given by $\Lambda_{jj} = 4\pi n \hbar \tilde{a}_j / m_j$, where $\tilde{a}$ is the background off-resonant *s*-wave scattering length and $m_j$ is the mass of a single particle for the *j*th species, respectively, while the nondiagonal terms are given by $\Lambda_{jk} = \Lambda_{kj} = 2\pi n \hbar \tilde{a}_{jk} / \mu_{jk}$, where $\tilde{a}_{jk}$ is the interspecies background off-resonant *s*-wave scattering length and $\mu_{jk} = m_j m_k / (m_j + m_k)$ are the reduced masses. The parameter $n$ denotes the mean density of particles: $n = N/V$, where $N$ is the number of "atomic particles" and $V$ is the volume of trapped particles (each molecule is being considered as two "atomic particles"), and $\hbar$ is Planck's constant divided by $2\pi$. In the case of Feshbach association of utracold bosonic atoms the atom-molecule coupling is given as $U = \sqrt{n} g / \hbar$, where $g = \hbar \sqrt{8\pi \tilde{a}_1 \Delta B \Delta \mu / m_1}$ [34,35]. In this expression $\Delta B$ is the width of the resonance, $\Delta \mu$ is the difference in magnetic momentum between the atomic and the bound molecular states. The detuning $\delta_t$ is given as $\delta_t = \Delta \mu [B(t) - B_0]/\hbar$, where $B(t)$ is external magnetic field, $B_0$ denotes the position of the Feshbach resonance. System (1) describes a lossless process, i.e., it preserves the total number of particles that we normalize to unity: $|a_1|^2 + 2|a_2|^2 = \text{const} = 1$. We consider the basic situation when the system starts from the all-atomic state: $|a_1(-\infty)| = 1$, $a_2(-\infty) = 0$. In the present paper we discuss the case of the LZ model hence hereafter we put $U(t) = U_0 = \text{const}$ and $\delta_t = 2\delta_0 t$.

It can be shown that the dynamics of the molecular state probability $p = |a_2|^2$ is described by the following nonlinear ordinary differential equation of third order:

$$p_{ttt} - \frac{G_t}{G} p_{tt} + \left[G^2 + 4\lambda(1-3p)\right] p_t + \frac{\lambda}{2} \frac{G_t}{G}(1 - 8p + 12p^2) = 0, \tag{2}$$

where
$$G = 2t - \Lambda_a + 2\Lambda_s p, \tag{3}$$

$$\Lambda_a = 2\Lambda_{11} - \Lambda_{21}, \quad \Lambda_s = \Lambda_a + \frac{1}{2}(\Lambda_{22} - 2\Lambda_{12}), \tag{4}$$

($\Lambda_{12} = \Lambda_{21}$) and $\lambda$ is the standard LZ parameter: $\lambda = U_0^2 / \delta_0$. In Eqs. (2)-(4) the independent variable and the parameters involved have been scaled as follows: $t' = \sqrt{\delta_0} t$ and $\Lambda'_{jk} = \Lambda_{jk} / \sqrt{\delta_0}$ ($j,k = 1,2$) and, for simplicity of notations, the primes have been omitted.

Note that the variation range of the function $p$ is $p \in [0, 1/2]$. However, since the quantity $Np(t)$ defines the number of molecules existing in the system at the point of time $t$, we conventionally refer to $a_2$ as to molecular sate probability amplitude, and to $p = |a_2|^2$ as to molecular sate probability.

If the cubic nonlinearities are not taken into account, i.e., if we put $\Lambda_{jk} = 0$ ($j, k = 1, 2$), then the function $G$ coincides with the LZ detuning $2t$. Hence, in a sense, the function $G$ plays the role of the effective (nonlinear) detuning and the point $t = t_{res}$ defined from the condition $G(t_{res}) = 0$ is the point of the effective resonance. Thus, we conclude that the introduction of the cubic nonlinearities results in a *nonlinear shift of the resonance*. Moreover, the structure of the effective detuning $G$ suggests that at sufficiently large absolute values of the variable $t$, when the condition $|2t| \gg |\Lambda_a - 2\Lambda_s p|$ holds, the role of the terms proportional to the parameter $\Lambda_s$ becomes negligible.

Further we notice that the parameter $\Lambda_a$ merely leads to a constant shift in the detuning which can be eliminated by the following change of the time variable: $t'' = t - \Lambda_a / 2$. This change does not affect the initial conditions since they are imposed at infinity ($t = -\infty$). Again, for simplicity of notation, we omit the double prime in what follows. [This is formally equivalent to removing the summand $\Lambda_a$ in Eq. (3)]. Hence the inter-particle elastic scattering is now described by a sole combined parameter $\Lambda_s$. As it can be seen from Eq. (2), there exist some nonzero parameters $\Lambda_{jk}$ for which the inter-particle elastic interactions merely result in the shift of the detuning by a constant which can be eliminated by the above mentioned change of the time variable. This occurs when the parameter $\Lambda_s$ is equal to zero.

We start our discussion by outlining some observations gained from numerical simulations. The dependence of the final transition probability to the molecular state $p(+\infty)$ on the parameters $\lambda$ and $\Lambda_s$ is shown in Fig. 1. As it is immediately seen, for a fixed $\Lambda_s$, the final transition probability is a monotonic function of $\lambda$ (see also Fig. 2a). Furthermore, $p(+\infty)$ is also a monotonic function of $\Lambda_s$ for fixed $\lambda$ (see Fig. 2b). This is an important conclusion gained from the 3-dimensional plot. Compared with the case when no inter-particle interactions are included ($\Lambda_s = 0$), the transition probability is always higher for negative $\Lambda_s$ and it is lower when $\Lambda_s$ is positive (Figs. 2a, 2b). Physically, this implies that atom-atom and molecule-molecule repulsive interactions diminish the molecule conversion efficiency while atom-molecule repulsion results in its increase. Thus, we conclude that the atom-atom, molecule-molecule attractive and atom-molecule repulsive interactions are favorable for molecule conversion efficiency. Time-dynamics of molecule formation also exhibits remarkable differences depending on whether the value of the parameter $\Lambda_s$ is negative or positive (see Fig. 3). Compared to the case when $\Lambda_s = 0$, at $\Lambda_s < 0$, the passage through the effective resonance occurs later, the transition to the molecular state takes place more slowly, and the amplitude and the frequency of the emerging oscillations are smaller. At $\Lambda_s > 0$ one observes the opposite behavior of these features. Hence, the general conclusion is that for the LZ model higher laser field intensities and large negative effective interactions $\Lambda_s$ are the favorable conditions for the formation of molecules.

Figure 3 also indicates that besides the time of the effective resonance crossing $t = t_{res}$, there exists another important time characterizing the association process – the point $t = t_{osc}$ at which the nonoscillatory evolution of the molecular state probability changes to an oscillatory behavior. Analyzing the system (1) from the point of view of classical Hamiltonian mechanics, one can see that the observed oscillations appear after the exact phase trajectory of the system crosses the separatrix in the phase space of the time-independent version of the system [27,28,31].

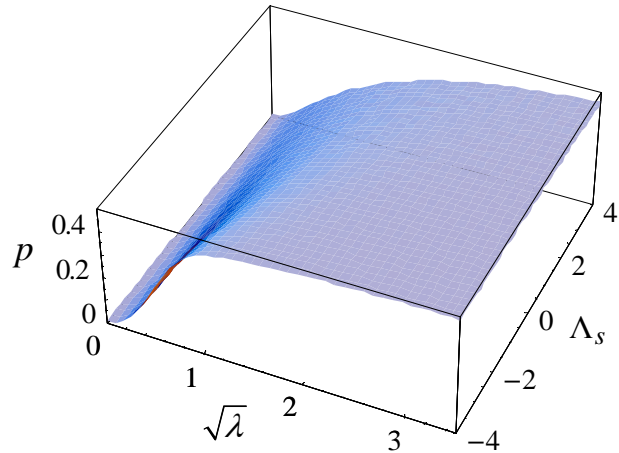

Fig. 1. Final transition probability to the molecular state versus $\sqrt{\lambda}$ and $\Lambda_s$. It is seen that the probability is a monotonic function of $\lambda$ for a fixed $\Lambda_s$ and it is also a monotonic function of $\Lambda_s$ for fixed $\lambda$.

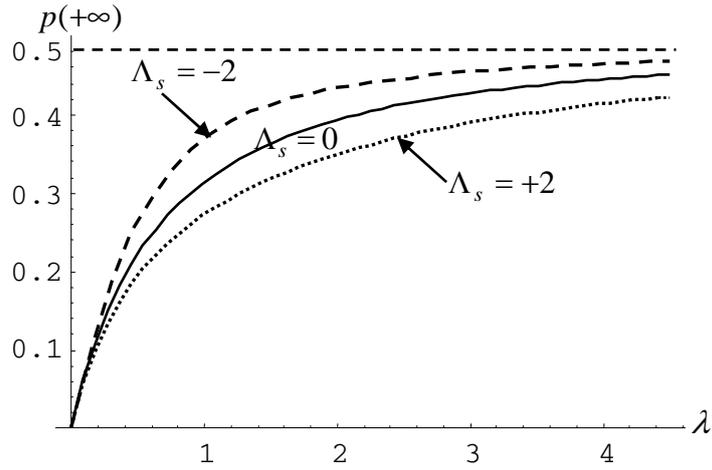

Fig.2a. Final transition probability to the molecular state versus $\lambda$ for different values of $\Lambda_s$.

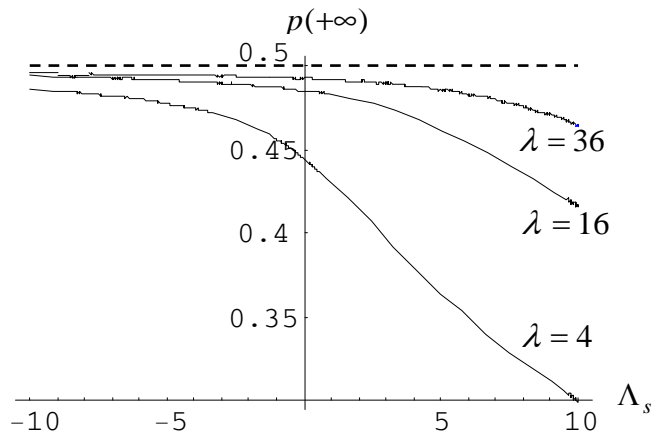

Fig.2b. Final transition probability to the molecular state versus $\Lambda_s$ for different values of $\lambda$.

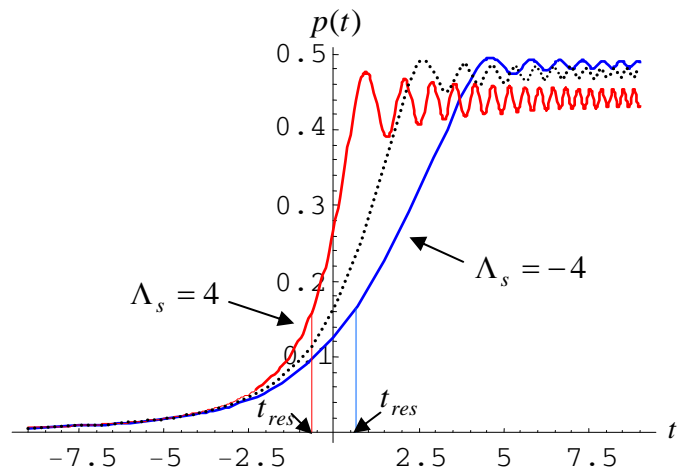

Fig. 3. The molecular state probability as a function of time at $\lambda = 9$. Dotted line corresponds to the case $\Lambda_s = 0$ while the solid lines correspond to the cases $\Lambda_s = +4$ and $\Lambda_s = -4$.

## 3. Mathematical treatment

To describe the presented features of the association process quantitatively, we proceed to the analysis of the equation for the molecular state probability (2). We consider *the strong nonlinearity regime* corresponding to high field intensities and we thus suppose that $\lambda$ is a large parameter. Since the function $G$ also adopts large values, we suppose that the leading terms in equation (2) are the last two. Hence, we make an attempt to construct an approximation by neglecting the two higher order derivative terms in the exact equation (2) and adding to the obtained truncated equation a term of the form $AG_t/G$:

$$\left[G^2 + 4\lambda(1-3p_0)\right]p_{0t} + \frac{\lambda}{2}\frac{G_t}{G}(1-8p_0+12p_0^2) - A\frac{G_t}{G} = 0, \tag{5}$$

where $A$ is a fitting parameter that will be specified afterwards. Applying the method presented in [36], we find the general solution to the limit equation (5):

$$\frac{1}{G^2(p_0)} = \frac{1}{9\lambda}\frac{p_0(p_0-\beta_1)(p_0-\beta_2)+C_0}{(p_0-\alpha_1)^2(p_0-\alpha_2)^2}, \tag{6}$$

where 
$$\alpha_{1,2} = \frac{1}{3} \mp \frac{1}{6}\sqrt{1+\frac{6A}{\lambda}}, \quad \beta_{1,2} = \frac{1}{2} \mp \sqrt{\frac{A}{2\lambda}} \tag{7}$$

and $C_0$ is the integration constant. This relation defines a *quintic algebraic equation* for the determination of the function $p_0(t)$. First of all, we note that the initial condition $p_0(-\infty)=0$ implies that $C_0=0$. Further, we see that at $t \to +\infty$ the left-hand side of Eq. (6) tends to zero and hence $p_0(+\infty)$ must be either $\beta_1$ or $\beta_2$. But since $\beta_2 > 1/2$ and the probability function $p_0$ cannot exceed $1/2$, we conclude that

$$p_0(+\infty) = \beta_1. \tag{8}$$

Thus, the approximate value of the final probability for the molecular state equals to $\beta_1$. Furthermore, one can determine a time $t=t_{res}$ such that $G(t_{res})=0$, i.e., a time at which the effective detuning $G$ passes through the effective resonance:

$$2t_{res} + 2\Lambda_s p_0(t_{res}) = 0. \tag{9}$$

From Eq. (6) it is clear that either $p_0(t_{res})=\alpha_1$ or $p_0(t_{res})=\alpha_2$. However, since $\alpha_2 > 1/2$, it must be

$$p_0(t_{res}) = \alpha_1. \tag{10}$$

Thus, the parameter $\alpha_1$ defines the approximate value of the molecular state probability at the effective resonance-crossing point. From Eqs. (9)-(10) it follows that

$$t_{res} = -\Lambda_s \alpha_1. \tag{11}$$

In order to develop general principles from which the fitting parameter $A$ can be determined, we insert the approximate solution $p_0(t, A)$ into the exact equation for the molecular state probability (2) and consider the behavior of the remainder

$$R = p_0''' - \frac{G_t}{G} p_0'' + A \frac{G_t}{G}. \tag{12}$$

It is intuitively clear that a better approximation $p_0$ should yield a smaller remainder [the latter would be identically zero if $p_0$ is the exact solution to Eq. (2)]. Thus, we try to minimize the remainder via appropriate choice of the fitting parameter $A$. We choose the fitting parameter $A$ by the condition that the remainder should not diverge at the effective resonance crossing $t_{res}$. This condition leads to the equation

$$p_0''(t_{res}) - A = 0. \tag{13}$$

The analysis of Eq. (13) then yields

$$A = \frac{4}{9\lambda} \text{Exp}\left[\left(1 - \frac{1}{2\lambda}\right) \frac{\Lambda_s}{\sqrt{\lambda}} + \frac{|\Lambda_s|}{2\pi} \sin\left(\frac{\pi \Lambda_s}{\lambda}\right)\right]. \tag{14}$$

If the condition $\Lambda_s \ll \sqrt{\lambda}$ holds then the following approximation can be used:

$$A = \frac{4}{9\lambda}\left(1 + \sqrt{\frac{2}{3}} \frac{\Lambda_s}{\sqrt{\lambda}}\right). \tag{15}$$

Comparison of the limit solution $p_0$ with the numerical solution shows that $p_0$ still misses several essential features of the association process (see Fig. 4). Indeed, for instance, the coherent oscillations between atomic and molecular populations which come up after the system passes through the resonance point are not contained in this approximation. The shortcomings of the limit solution $p_0$ are caused by the singular procedure used to obtain it. Indeed, we have constructed $p_0$ by neglecting the two highest order derivative terms in Eq. (2). Of course, when determining the optimal value of $A$ we have afterwards taken into account these terms, to some extent.

To improve the result, we need a next correction term that takes into account the second and third order derivatives of $p$. However, it turns out that this is not a simple task because the equation obeyed by the exact correction term $u \equiv p - p_0$ is still an essentially non-linear one.

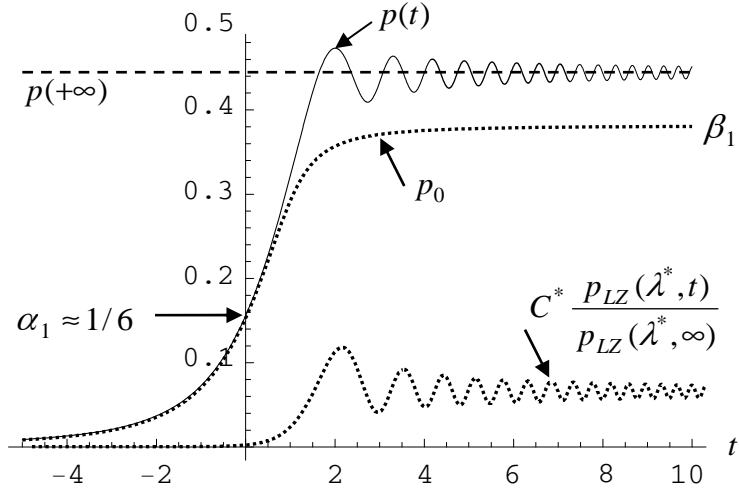

Fig. 4. Molecular state probability $p(t)$, the limit solution $p_0$ determined from Eq. (6), and the scaled solution to the linear LZ problem with modified parameters [Eq. (18)].

To develop an appropriate approach, we first consider the LZ crossing in the relatively simple case when the cubic nonlinearities are neglected, i.e., we take $\Lambda_s = 0$. Now, by introducing in Eq. (2) the change of dependent variable

$$p = p_0 + u, \tag{16}$$

we obtain an exact nonlinear differential equation for the correction $u$ which we write in the following factorized form:

$$\left(\frac{d}{dt} - \frac{1}{t}\right)\left(u'' + 4\left[t^2 + \lambda(1 - 3p_0)\right]u + p_0'' - A - 6\lambda u^2\right) - 4tu = 0. \tag{17}$$

Since the function $p_0$ is already a good first approximation, the correction $u$ is supposed to be small. Further we notice that if in (17) we neglect the nonlinear term $-6\lambda u^2$ and consider $p_0$ as a constant then the solution of the equation can be written as a scaled solution to the linear LZ problem [1] with a modified LZ parameter. This observation gives an argument to make the conjecture that the exact solution of Eq. (17) can be approximated as

$$u = C^* \frac{P_{LZ}(\lambda^*, t)}{P_{LZ}(\lambda^*, \infty)}, \tag{18}$$

where $P_{LZ}(\lambda^*, t)$ is the solution of the linear LZ model [1] which can be expressed in terms of confluent hypergeometric functions [37], and $C^*$ and $\lambda^*$ are fitting parameters which will be determined afterwards. This conjecture is well confirmed by numerical analysis; the numerical simulations show that one can always find $C^*$ and $\lambda^*$ such that the function (18) accurately fits the numerical solution to the exact equation (17).

To obtain analytical expressions for the fitting parameters $C^*$ and $\lambda^*$, we substitute the trial function (18) into the exact equation (17) and aim at minimization of the remainder

$$R = \left(\frac{d}{dt} - \frac{1}{t}\right)\left\{4[\lambda(1-3p_0) - \lambda^*]\frac{P_{LZ}(\lambda^*,t)}{P_{LZ}(\lambda^*,\infty)} + \frac{2\lambda^*}{P_{LZ}(\lambda^*,\infty)} + \frac{1}{C^*}(p_0'' - A) - 6\lambda C^* \frac{P_{LZ}^2(\lambda^*,t)}{P_{LZ}^2(\lambda^*,\infty)}\right\} \quad (19)$$

via appropriate choice of $C^*$ and $\lambda^*$.

The analysis of the behavior of the first term in the curly brackets suggests that the remainder is strongly suppressed if one chooses

$$\lambda^* = \lambda(1 - 3p_0(+\infty)). \quad (20)$$

Taking into account the value of $p_0(+\infty)$ [defined by Eq. (8)], we rewrite Eq. (20) as follows:

$$\lambda^* = -\frac{\lambda}{2} + 3\sqrt{\frac{A\lambda}{2}}. \quad (21)$$

Hence, for $\lambda \gg 1$, $\lambda^*$ is a large *negative* parameter. This choice of $\lambda^*$ leads to an important observation. It is known that [1]

$$\lim_{t \to +\infty} P_{LZ}(\lambda,t) = 1 - e^{-\pi\lambda}, \quad (22)$$

hence, in the case of negative $\lambda^*$ the function $P_{LZ}(\lambda^*,\infty)$ grows exponentially with $|\lambda^*|$. Consequently, for this choice of $\lambda^*$ the second term in the curly brackets in Eq. (19) is also essentially suppressed. Regarding the two last terms in Eq. (19), one should minimize them with respect to the parameter $C^*$. This implies the condition

$$\frac{\partial(R/C^*)}{\partial C^*} = \left(\frac{d}{dt} - \frac{1}{t}\right)\left(-\frac{1}{C^{*2}}(p_0'' - A) - 6\lambda \frac{P_{LZ}^2(\lambda^*,t)}{P_{LZ}^2(\lambda^*,\infty)}\right) = 0. \quad (23)$$

Since the last term is proportional to (large) $\lambda$ and $P_{LZ}(\lambda^*,t)$ is an increasing function of time the "worst" point is $t = +\infty$. Hence, we look for minimization at $t = +\infty$. This immediately leads to the following value for $C^*$:

$$C^* = \sqrt{\frac{A}{6\lambda}}. \quad (24)$$

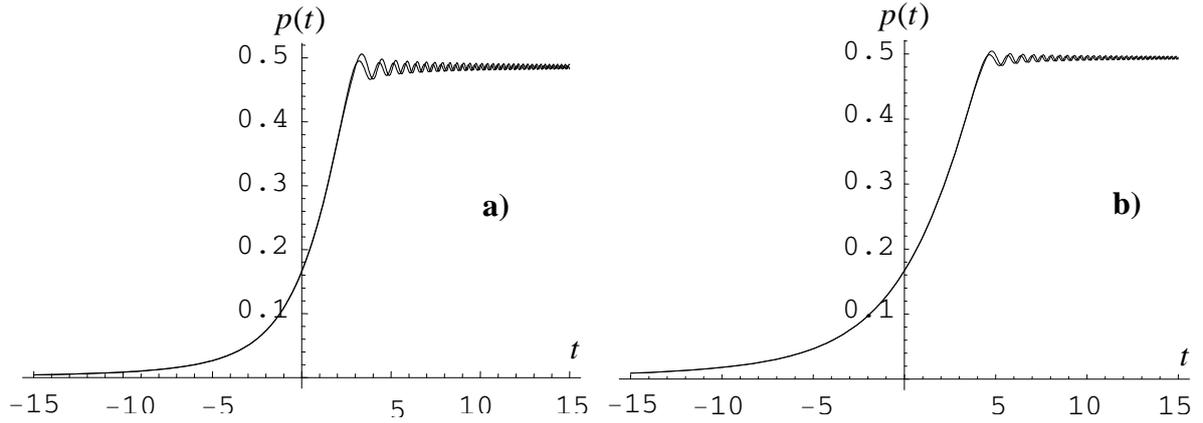

Fig. 5. Molecular state probability $p(t)$ and the approximate solution given by Eqs. (16) and (18) as functions of time for a) $\lambda = 15$ and b) $\lambda = 36$. The fitting parameters are taken as $A = 4/(9\lambda)$, $C^* = \sqrt{A/(6\lambda)}$, and $\lambda^* = -\lambda/2$. The analytical formula slightly overestimates the final transition probability.

The comparison of the constructed approximation with the numerical solution shows that formulas (21) and (24) define a quite good approximation which describes the dynamics of the system qualitatively well (see Fig. 5). Taking into account Eqs.(8), (16), and (18), it can easily be seen that the final ($t \to +\infty$) transition probability to the molecular state is given by the following relation:

$$p(+\infty) = \frac{1}{2} - \sqrt{\frac{A}{2\lambda}} + C^*. \qquad (25)$$

This relation shows that the final transition probability does not depend on parameter $\lambda^*$. Obviously, it is changed with variation of $A$ and $C^*$ (note that variation of $A$ inevitably leads to variation of $C^*$). By analyzing the structure of the constructed approximate equation [see Eqs. (16) and (18)], we see that the first term of the constructed two-term solution is a step-wise function while the second one describes the oscillations which come up after the system has passed through the resonance (see Fig. 4). The frequency of these oscillations is defined by the value of the parameter $\lambda^*$ only. Variation of the parameter $C^*$ is not potent to change the frequency of the oscillations since $C^*$ is just the scaling parameter in Eq. (18). Summing up these observations we arrive at a conclusion that the introduced parameters $\lambda^*$ and $C^*$ characterize qualitatively different physical processes; the parameter $C^*$ describes the final transition probability to the molecular state, whereas the parameter $\lambda^*$ determines the frequency of the oscillations, emerging some time after the system has passed through the resonance. Though to construct an approximate solution we use a solution of a linear equation $P_{LZ}(\lambda^*,t)$, the parameters involved in the constructed approximation (18), $\lambda^*$ and $C^*$, are essentially determined by the nonlinear terms involved. Note that the values of the parameters $\lambda^*$ and $C^*$ depend on the value of the fitting parameter $A$.

Analytical expressions (21) and (24) have been obtained when attempting to suppress the remainder (19) as much as possible. However, from the mathematical point of view, to obtain an accurate approximation, one should minimize the next approximation term $w = p - p_0 - u$ and not the remainder itself. It can be seen that the remainder (19) serves as the inhomogeneous term of the exact equation obeyed by $w$. Thus, we try to minimize the next approximation term $w$ via appropriate variation of the remainder. By applying the described approach we arrive at a conclusion that the result given by Eqs. (21) and (24) can be considerably improved if we redefine the fitting parameters as follows:

$$C^* = \sqrt{\frac{A}{6\lambda} - \frac{1}{54\lambda}} \quad \text{and} \quad \lambda^* = \lambda(1 - 3[p_0(+\infty) + C^*]). \tag{26}$$

The comparison of the refined approximation with the numerical solution shows that it is a very good approximation at $\lambda > 2$.

Now, we return to the general case with $\Lambda_s \neq 0$. Based on the experience gained for $\Lambda_s = 0$, we make the conjecture that the approximate solution in this general case has an analogous structure:

$$p = p_0 + C^* \frac{P_{LZ}(\lambda^*, t - t_{ph})}{P_{LZ}(\lambda^*, \infty)}, \tag{27}$$

where the parameters $\lambda^*$ and $C^*$ are still defined by formula (26) and $t_{ph}$ is the newly introduced fitting parameter. Eq. (27) along with expressions (14) and (26) for the involved fitting parameters is the main result of the present paper. The first summand of Eq. (27), $p_0$, is a step-wise function while the second one monotonically increases until the small-amplitude oscillations appear (see Fig. 4). When presenting general observations, we have already mentioned that inter-particle elastic scattering results in the shift of both effective resonance point, $t = t_{res}$, and the point where the small-amplitude oscillations start, $t = t_{osc}$, as compared to the case when the inter-particle elastic scattering is neglected. Hence, the fitting parameter $t_{ph}$ introduced in the approximation (27) is supposed to describe the shift of the point where the small-amplitude oscillations start. Supposing that the fitting parameter $t_{ph}$ is related to the effective resonance crossing point $t_{ph} = t_{ph}(t_{res})$ we further try to derive an analytical expression for this dependence. To this end, assuming that $t_{ph}$ is proportional to $t_{res}$, we determine the coefficient of proportionality numerically: $t_{ph} \approx 2.8 t_{res}$. The physical processes emerging due to inter-particle scattering are described via the dependence of the parameters $A$ and $t_{res}$ on $\Lambda_s$. Comparison of the approximation (27) with the numerical solution shows that it is a very good approximation for $\lambda > 2$ and $-0.5 \leq \Lambda_s / \sqrt{\lambda} \leq 0.25$; it accurately describes the association process for almost all the time range.

To analyze the behavior of the final transition probability, we substitute the values of the fitting parameters $A$, $\lambda^*$ and $C^*$ determined by Eqs. (14) and (26) into expression for the final probability of transition to the molecular state (25). This results in the following relation:

$$p(+\infty) = \frac{1}{2} - \frac{1}{\lambda}\left(\left[\frac{\sqrt{2}}{3} - \frac{\sqrt{6}}{9}\right]e^{\gamma/2} + \frac{1}{54}\right) \approx \frac{1}{2} - \frac{1}{\lambda}\left(0.1992 e^{\gamma/2} + 0.0185\right), \qquad (28)$$

where
$$\gamma = \left(1 - \frac{1}{2\lambda}\right)\frac{\Lambda_s}{\sqrt{\lambda}} + \frac{|\Lambda_s|}{2\pi}\sin\left(\frac{\pi \Lambda_s}{\lambda}\right). \qquad (29)$$

Formula (28) is one of the most relevant results of the present paper. This formula (28) agrees well with the results of numerical simulations (Fig. 6); it also confirms the statement that negative effective scattering $\Lambda_s < 0$ is favorable for molecule formation (within the applicability range of the formula). Indeed, if $\Lambda_s < 0$ then $\gamma < 0$, hence, the final transition probability increases. Obviously, when $\Lambda_s > 0$ the final transition probability decreases. The maximum discrepancy between numerical and analytical solutions shown in Fig. 6 corresponds to $\lambda = 5$, $\Lambda_s = 0.7$ and equals $0.001540$. In the case $\Lambda_s = 0$ expression (28) takes the following form:

$$p(+\infty) = \frac{1}{2} - \frac{1}{\lambda}\left(\frac{\sqrt{2}}{3} + \frac{1}{54} - \frac{\sqrt{6}}{9}\right) \approx \frac{1}{2} - \frac{0.2178}{\lambda}. \qquad (30)$$

This formula confirms the result of Refs. [17-28] stating that in the strong coupling limit, the final probability for non-transition to the molecular state is inversely proportional to the Landau–Zener parameter (in contrast to the linear two-state case when the dependence is exponential [1]).

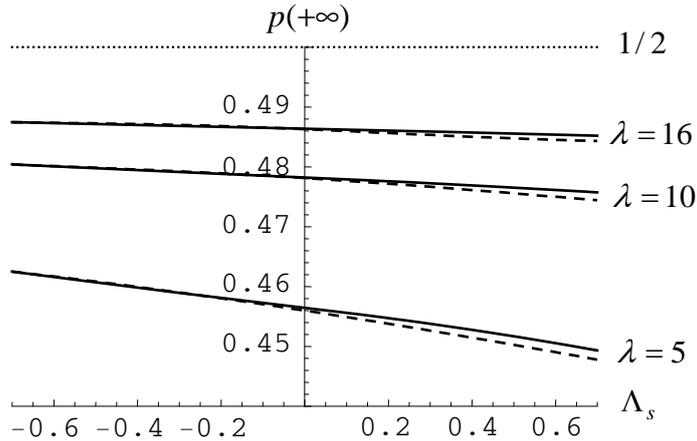

Fig. 6. Final transition probability versus $\Lambda_s$ for $\lambda = 5$, $\lambda = 10$, and $\lambda = 16$.
Solid line - analytical solution (28), dashed line – numerical solution.

The method we apply in the present paper to tackle the problem is analogous to that presented in Refs [19,20], where the basic nonlinear version of the NLZ problem has been considered. In these papers the inter-particle elastic scattering has not been taken into account. It has been shown that the approximate solution to the problem can be written as a sum of two distinct terms, a solution of a limit first-order nonlinear equation and a scaled solution of the linear Landau-Zener problem with modified parameters. In this case the solution of the limit equation has been shown to be determined as a solution of a polynomial equation of *fourth* order. However, as we have seen above, inclusion of the cubic-nonlinear terms describing inter-particle elastic scattering results in modification of the limit equation [see Eq. (5)]: now, the solution of this equation is given as a solution of a polynomial equation of *fifth* order (6). Note that if we put $\Lambda_s = 0$ the polynomial equation of fifth order will reduce to a polynomial equation of fourth order used in Refs. [19,20].

Finally, we would like to mention that the physical situation we have been discussing is realized under current experiments (for a comprehensive review see Ref. [38]). A typical example is the 85Rb experiment performed by Hodby and co-workers in JILA [39], where coherent formation of Rb2 molecules via sweep of the magnetic field through the Feshbach resonance located at $155\,G$ is realized. The magnetic field is changed at a given linear sweep rate $\dot{B}$, and the molecule conversion efficiency is measured as a function of the inverse sweep rate. Thus, the external field configuration applied in this experiment corresponds to the LZ model. The initial density of the atomic cloud $n$ is of the order of $10^{11}\,\text{cm}^{-3}$, the background scattering length of atoms is $\tilde{a}_1 = -443 a_0$, where $a_0$ is the Bohr radius, the resonance width is $\Delta B = 10.71\, G$, the difference in magnetic momentum between the atomic and the bound molecular channels is $\Delta \mu = -2.33 \mu_B$, where $\mu_B$ is the Bohr magneton. The LZ parameter is written as $\lambda = 16\pi n \hbar \tilde{a}_1 \Delta B/(\dot{B} m_1)$. At small enough sweep rates and high enough atomic densities applied at this experiment the molecule formation is described by the strong interaction regime $\lambda \gg 1$ discussed here; indeed, for the sweep rate $1/\dot{B} = 1000 \mu s/G$ and $n = 5 \cdot 10^{11}\, cm^{-3}$ one has $\lambda \approx 5$. Furthermore, estimating the value of the dimensionless parameter $\Lambda_s$, we see that in this particular experiment $\Lambda_s \approx 10^{-2} \ll \lambda$. Hence, the presented theory is helpful for interpretation of the mentioned experiment.

## 4. Summary

We have presented a nonlinear version of the LZ problem that arises in the theory of coherent photoassociation or Feshbach resonances in atomic Bose-Einstein condensates, focusing on the role of the atom-atom, atom-molecule, and molecule-molecule scattering which are described by the cubic nonlinear terms in the system (1). We have shown that the interparticle interactions strongly affect the dynamics of the molecule formation in the vicinity of the resonance, resulting in the nonlinear shift of the resonance point [see Eqs. (3)-(4)]. We have proven that in the case of the LZ model the inter-particle elastic scattering is described by a sole combined parameter $\Lambda_s$ (this fact has already been noticed in Ref. [32]). By studying both final ($t \to +\infty$) transition probability to the molecular state and temporal dynamics of molecule formation, we have arrived at a general conclusion that for the LZ large values of the LZ parameter $\lambda$ and large negative effective interactions $\Lambda_s$ are the most favorable conditions for the formation of molecules.

Further, we have undertaken a variational treatment to the NLZ problem in the strong coupling limit. Using the third-order nonlinear differential equation for the molecular state probability (2), we have constructed an approximate solution to the problem in three steps.

1. Neglecting two higher order derivative terms in the exact equation for the molecular state probability (2), we define the nonlinear limit equation (5) in which we introduce an adjustable parameter $A$. We explicitly solve the limit equation (5) and further determine $A$ from the condition of minimization of the remainder (19). Note that the obtained value of $A$ depends on $\Lambda_s$.

2. Then, we consider the case $\Lambda_s = 0$. We insert $p = p_0 + u$, into the exact equation (2) and make a conjecture that the correction $u$ can be represented as a scaled solution of the linear LZ problem, containing some effective LZ parameter $\lambda^*$ [see Eq. (18)]. Again, the fitting parameters $\lambda^*$ and $C^*$ are determined via minimization of the remainder (19). This defines $\lambda^*$ and $C^*$ in terms of the parameter $A$ [see Eq. (26)].

3. To construct an appropriate approximation in the general case when $\Lambda_s \neq 0$, we make a conjecture that in this case the approximate solution has the same structure as for the case $\Lambda_s = 0$ and the parameters $C^*$ and $\lambda^*$ are still determined from Eq. (26) but now the function (27) takes into account the interparticle elastic scattering due to the dependence of the parameter $A$ on $\Lambda_s$ and the introduced shift in the argument of the function.

The described approach can be viewed as a variational method. It enables one to construct a highly accurate and simple analytic approximation describing the time dynamics of the coupled atom-molecular system at $\lambda > 2$ and $-0.5 \leq \Lambda_s / \sqrt{\lambda} \leq 0.25$ (Fig. 6). Moreover, the decomposition (27) shows that the solution can be separated into two distinct parts: $p_0$, describing the process of molecule formation, and $u$, describing the remaining oscillations which come up after the system has passed through the effective resonance. This decomposition clearly indicates that the process of molecule formation is mainly governed by the nonlinear limit equation (5). It should be stressed that the derived approximate solution for the first time describes the whole temporal dynamics of the nonlinear LZ problem with inter-particle elastic interactions included.

Finally, we note that the presented approach is not restricted to the particular LZ problem treated here. It can be easily generalized to other time-dependent models. Hence, the developed method is a general strategy for attacking analogous nonlinear two-state problems.

**Acknowledgments**

A. Ishkhanyan acknowledges the support from the Armenian National Science and Education Fund (ANSEF Grant No. 2009-PS-1692) and the International Science and Technology Center (ISTC Grant No. A-1241). R. Sokhoyan acknowledges the support from INTAS (Young Scientist Fellowship Ref. No. 06-1000014-6484) and the French Embassy in Armenia (Grant No. 2006-4638 Boursière du Gouvernement Français). A. Ishkhanyan acknowledges Institut Carnot de l'Université de Bourgogne for the invited professorship in 2007. A. Ishkhanyan and R. Sokhoyan thank the Department of Physics and Astronomy, University of Turku (Finland) for kind hospitality. K.-A. Suominen acknowledges the financial support by the Academy of Finland project 115982.